\documentclass[a4paper,11pt]{article}
\usepackage{hyperref}
\usepackage[english]{babel}
\usepackage[utf8]{inputenc}
\usepackage{graphicx}
\usepackage{geometry}
\usepackage{setspace}

\begin{document}

\begin{titlepage}
    \begin{center}
        \includegraphics[width=\textwidth]{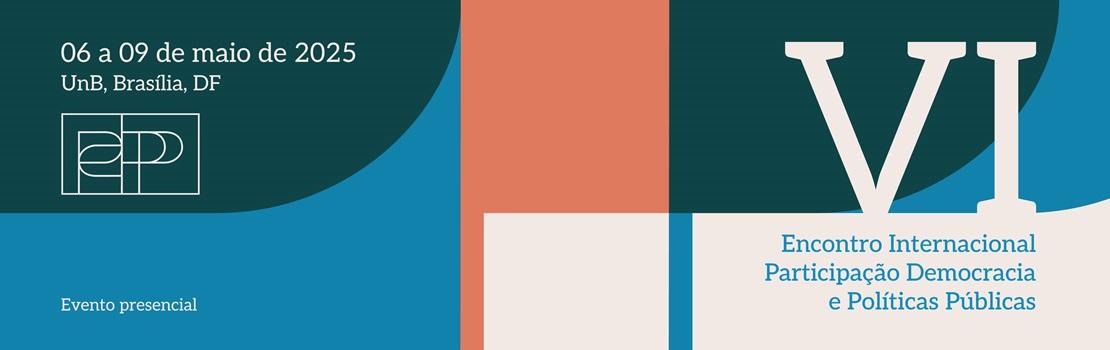}

        \vspace{1cm}

        {\Large VI Encontro Internacional Participação, Democracia e Políticas Públicas}\\[0.5cm]
        {\large 06 a 09/05/2025, UnB, Brasília (DF)}\\[2cm]

        {\large ST05. Instituições participativas e políticas públicas: legados, reconstrução e transformação}\\[1cm]

        {\Huge \textbf{Generative AI as a catalyst for democratic Innovation: Enhancing citizen engagement in participatory budgeting}}\\[2cm]

        {\large 	Ítalo Alberto do Nascimento Sousa; Jorge Machado;  José Carlos Vaz}\\
        {\small Universidade de São Paulo}
    \end{center}
\end{titlepage}

\title{Generative AI as a Catalyst for Democratic Innovation: Enhancing Citizen Engagement in Participatory Budgeting}
\date{}
\maketitle

\begin{abstract}
This research examines the role of Generative Artificial Intelligence (AI) in enhancing citizen engagement in participatory budgeting. In response to challenges like declining civic participation and increased societal polarization, the study explores how online political participation can strengthen democracy and promote social equity. By integrating Generative AI into public consultation platforms, the research aims to improve citizen proposal formulation and foster effective dialogue between citizens and government. It assesses the capacities governments need to implement AI-enhanced participatory tools, considering technological dependencies and vulnerabilities. Analyzing technological structures, actors, interests, and strategies, the study contributes to understanding how technological advancements can reshape participatory institutions to better facilitate citizen involvement. Ultimately, the research highlights how Generative AI can transform participatory institutions, promoting inclusive, democratic engagement and empowering citizens.


\end{abstract}

\section{Introduction}
\label{sec:intro}

In recent years, democratic governance has entered a phase of experimentation and innovation aimed at deepening citizen participation in public decision-making. Participatory Budgeting (PB), a process that empowers citizens to directly decide on the allocation of public funds, has emerged as a flagship example of this democratic innovation, spreading to thousands of cities worldwide~\cite{brennancenter1}.

At the same time, rapid advances in Generative Artificial Intelligence (AI) are opening new possibilities to enhance civic engagement. Generative AI systems, such as large language models, can produce human-like text, offering novel tools to support communication and creativity in governance processes.

This article explores how generative AI can serve as a catalyst for democratic innovation by enhancing citizen engagement in participatory budgeting, combining theoretical insights and practical analysis. Particular emphasis is placed on the Brazilian context, a country with a strong legacy of participatory budgeting, and especially on the case of the \textit{Participe+} platform, used by the S\~ao Paulo City Hall (\textit{Prefeitura de S\~ao Paulo}) for online participation.

The central question is whether integrating generative AI into participatory budgeting can improve the quality and inclusiveness of citizen proposals, foster greater public participation, and ultimately strengthen the legitimacy of policymaking. Brazil provides a rich setting for this investigation. The country was the birthplace of participatory budgeting in 1989 (Porto Alegre’s famed experiment) and has since institutionalized numerous participatory forums and digital platforms~\cite{archiveepa, brennancenter1}.

S\~ao Paulo’s \textit{Participe+} platform, launched in 2020 amid the COVID-19 pandemic, exemplifies the reconstruction of participatory institutions through digital means, achieving a dramatic expansion of participation in the city’s budgeting process~\cite{opengov1}. Now, as generative AI technologies mature, there is an opportunity to further transform these participatory processes by integrating AI assistance into how citizens formulate and discuss their budget proposals.

This article is structured to examine the intersection of participatory democracy and AI step by step. We begin with a theoretical background on participatory democracy, participatory budgeting, and digital governance. Next, a review of the literature surveys international and Brazilian studies on citizen participation, online engagement, and PB outcomes. We then present a case study of S\~ao Paulo’s \textit{Participe+} platform and its use in PB.

Building on this, we introduce generative AI, its definition, characteristics, and current applications in public participation, and explore how it could be applied specifically to proposal development in participatory budgeting. A conceptual framework is proposed to illustrate the integration of generative AI into each stage of the PB process, from data collection and proposal generation to feedback and interaction with public officials.

The methodology for our analysis is explained, combining a bibliographic review with insights gleaned from interactions with \textit{Participe+} coordinators and municipal managers. In the results and discussion section, we consider the expected outcomes of using AI in PB, including potential benefits for citizen engagement and process legitimacy, as well as the challenges and risks such an approach entails.

Finally, the conclusion reflects on whether AI-assisted participatory budgeting constitutes a genuine democratic innovation. It also discusses risks such as algorithmic bias, digital exclusion, and ethical concerns and suggests directions for future research and practice. In doing so, we relate our findings to the broader theme of participatory institutions and public policies: legacy, reconstruction, and transformation, arguing that generative AI could help reinvigorate the legacy of participatory budgeting by reconstructing citizen engagement processes in transformative ways.

\section{Participatory Democracy and Digital Innovation}

\subsection{Participatory Democracy}

Participatory democracy refers to a model of democracy in which citizens have meaningful opportunities to directly engage in deliberation and decision making, rather than relying solely on elected representatives. Classic democratic theorists have long argued that democracy should extend beyond periodic voting to include active citizen participation in governance.

Carole Pateman’s seminal work on participatory democracy (1970) emphasized that genuine democratic participation can educate and empower citizens, leading to more informed public decision-making~\cite{pateman1970}. In this theory, citizens and institutions are viewed as interdependent: robust participation can improve both individual civic skills and the quality of collective outcomes.

Although participatory democracy is often rooted in local, face-to-face settings (e.g., town hall meetings, neighborhood councils), its principles have been increasingly institutionalized in structured governance programs. The late twentieth century saw a wave of democratic innovations: new institutions designed to increase direct citizen input, such as deliberative forums, citizen assemblies, and participatory budgeting.

These innovations aim to complement representative democracy by making it more inclusive and responsive~\cite{berliner2024}. Political theorist Hélène Landemore, for example, argues that emerging technologies can help scale up participatory and deliberative democracy. However, the success of participatory democracy depends not only on tools and processes but also on power dynamics: If public inputs are ignored, participation risks becoming a mere symbolic 'window dressing'. Thus, the theory underscores both the promise of stronger citizen voice and the importance of government response.

\subsection{Participatory Budgeting}

Participatory budgeting is a concrete application of participatory democracy principles in the field of public finance decision making. PB is 'a democratic process in which community members decide how to spend part of a public budget'~\cite{pbp2023, Craveiro2016}. In typical PB processes, residents propose ideas for public projects, deliberate on priorities, and vote on which proposals should receive funding.

The concept originated in Porto Alegre, Brazil, in 1989, as part of a reformist effort by the Workers’ Party to promote transparency and accountability in a context of inequality and clientelism~\cite{brennancenter1}. The model involved neighborhood assemblies and elected citizen delegates who helped shape budget priorities. It led to substantial improvements in public services, such as expanded access to sanitation and clean water~\cite{archiveepa}.

During the 1990s, PB spread rapidly across Brazil. In the early 2000s, more than 140 municipalities (approximately 2.5\% of all Brazilian cities) had adopted the model~\cite{archiveepa}. Within two decades, nearly half of Brazil’s 250 largest cities had implemented PB programs, transferring on average 5 to 15\% of municipal budgets, and in some cases all of their capital investment funds, to citizen-selected projects~\cite{brennancenter1}. Studies found that Brazilian cities with PB experienced significantly larger reductions in infant mortality and poverty rates than comparable cities without it~\cite{brennancenter1}.

Despite these achievements, there are limitations. In some cases, the poorest residents remain underrepresented in PB processes~\cite{archiveepa}, and without sustained political will, the impact of PB on certain public services may be limited~\cite{brennancenter1}. However, PB has been recognized by the United Nations as a best practice in democratic governance~\cite{brennancenter1} and has spread to more than 7,000 cities around the world by the 2020s. This global expansion shows the adaptability of PB, but also its variability: Processes differ in scale, citizen power, and results. Still, the core idea remains: ordinary people directly influence public spending, institutionalizing democratic participation in government.

\subsection{Digital Governance and E-Participation}

The rise of digital technologies has opened new avenues for participatory democracy, leading to the concept of digital governance, or e-Governance, the use of information and communication technologies to support government operations and enhance citizen engagement.

Digital governance includes open data portals, online service delivery, and electronic participation platforms that allow citizens to consult, deliberate, and vote online~\cite{opengov1}. Integrating digital tools into participatory processes helps overcome time and geography barriers, making participation more accessible to people who cannot attend in-person meetings.

Early examples of digital participation include electronic town halls in the 1990s and online public policy forums. In the PB context, some cities began experimenting with online idea submission and voting in the 2000s. For example, Belo Horizonte launched a digital PB process in 2006, allowing residents to propose and vote for projects over the Internet~\cite{sampaio2011}. Although the deliberative quality of the discussions was modest, the studies showed that the participants engaged respectfully and provided reasons for their views.

Theoretical perspectives on digital participation emphasize both opportunities and risks. On the one hand, online platforms can scale participation to involve tens or hundreds of thousands of people, lowering barriers to entry (e.g., geographical or socioeconomic). On the other hand, digital divides (e.g., lack of access or digital literacy) can exclude some groups, and unmoderated forums can suffer from incivility or misinformation.

Increasing numbers of cities have adopted open source platforms like CONSUL (from Madrid Decide Madrid) and Decidim (developed in Barcelona), including São Paulo’s \textit{Participe+}~\cite{opengov1} that is based in CONSUL. These platforms offer tools for proposal submission, forums, voting, and transparency dashboards, representing a new generation of participatory infrastructure.

The evolution from e-government (focused on service efficiency) to e-democracy (focused on citizen empowerment) frames this trend. Generative AI, when integrated into these systems, can be seen as part of this evolution, potentially improving accessibility, scale, and usability of citizen participation tools~\cite{berliner2024}.

In summary, participatory democracy provides the normative foundation valuing direct citizen voice; participatory budgeting is a concrete institutional mechanism realizing that vision in budgeting; and digital governance offers new means to implement such participatory mechanisms at scale. These elements form the context in which the potential impact of generative AI must be understood: any AI integration should reinforce the democratic values of inclusion, deliberation, and empowerment that are central to the original ethos of participatory budgeting, while addressing the practical needs of scaling and improving participation in a digital era.

\section{Online Participation and Participatory Budgeting}

\subsection{International Perspectives on Citizen Participation and Online PB}

A rich body of international literature examines how citizen participation in governance affects public policy and community outcomes. Numerous case studies and comparative works on participatory budgeting highlight its benefits in diverse contexts. Experiences in Europe (e.g. Spain, France, Portugal), North America, and Asia have shown that PB can increase civic engagement and surface local priorities that may be overlooked by officials~\cite{brennancenter1}.

In terms of civic engagement, interviews with PB practitioners in the United States and Canada suggest that PB often boosts the sense of civic agency of citizens and strengthens community bonds~\cite{brennancenter1}, it demystifies municipal budgeting and fosters constructive interaction between residents and government. PB is also recognized for identifying unmet community needs, particularly in marginalized areas that top-down budgeting may neglect~\cite{brennancenter1}.

Quantitative studies have found that PB can lead to improved social outcomes. As noted previously, Brazilian municipalities with PB experienced significant reductions in infant mortality~\cite{brennancenter1}.  Similar participatory frameworks applied in other countries have yielded positive outcomes in sectors such as health and sanitation. In the long term, a better understanding of the budget can improve public engagement and
relations between governments and citizens~\cite{Craveiro2015}. However, international research also cautions that results depend heavily on process design and implementation context.

For example, a study in Peru found no significant improvements in the water services of PB, indicating that institutional and resource restrictions can undermine its effectiveness~\cite{brennancenter1}. In Indonesia, researchers observed that low-income neighborhoods benefited less from PB, as poorer residents participated at lower rates, highlighting persistent socioeconomic barriers, even in 'open' processes~\cite{brennancenter1}.

To maximize equity, PB processes must intentionally include underrepresented groups, employing outreach, support mechanisms, and possibly technology-based assistance, an opening for AI interventions. Comparative research also highlights differences in the scale and authority of PB programs. Although many Latin American PBs are citywide with substantial budget allocations, PBs in North America or Europe are often limited to districts or neighborhoods, with smaller funds~\cite{brennancenter1}. These structural differences can affect citizen motivation and participation.

The rise of online participation has further expanded the PB landscape. Early experiments, such as Belo Horizonte’s digital PB in Brazil, demonstrated that online platforms could facilitate deliberative exchanges, although often with lower levels of back-and-forth interaction compared to in-person assemblies~\cite{sampaio2011}. As technology matured, platforms like Madrid’s \textit{Decide Madrid} (launched in 2015) attracted tens of thousands of users who proposed and voted on public initiatives, provided the interface was accessible and well publicized~\cite{veciana2023}.

Researchers studying these platforms found that while the volume of participation increased dramatically, it also brought challenges. Cities experienced an overload of information from thousands of proposals and comments, making it difficult for both public officials and citizens to meaningfully process input~\cite{veciana2023}. In response, cities like Hamburg have developed natural language processing (NLP) tools to help analyze citizen contributions. Hamburg’s DIPAS system, for example, used clustering and topic extraction to synthesize public feedback into digestible themes~\cite{veciana2023}. One key lesson from that project was the importance of maintaining links between AI-generated summaries and original citizen statements, to preserve trust and context~\cite{veciana2023}.

Another key theme in the literature is the quality of online discourse. Deliberation scholars, including Sampaio et al.~\cite{sampaio2011}, argue that even when reciprocity in online forums is limited, there is civic value in articulating arguments and reading others' proposals, which contributes to democratic learning. However, concerns remain about superficial engagement, echo chambers, or click-based voting. Research agrees that platform design plays a crucial role: features such as discussion prompts, justification requirements, and moderation mechanisms influence the deliberative quality of participation.

Security and authenticity are also major concerns in digital PB. Risks include fraud, bots, and duplicate submissions. Some countries have addressed this through secure digital identification systems. For example, Brazil’s national participatory platform linked submissions to the government's unified log-in system to ensure unique user identities~\cite{veciana2023}. During the 2023 Participatory Pluriannual Plan process, this system prevented malicious attacks and minimized hate speech, with only a small fraction of the 8,200+ proposals requiring moderation~\cite{veciana2023}.

This example illustrates that online processes can maintain procedural integrity, an essential condition for public trust.

\subsection{Brazilian Studies and Data on Participation and Participatory Budgeting}

Brazil’s experience with participatory institutions has been extensively studied due to its pioneering role in participatory democracy. Scholars such as Leonardo Avritzer and Brian Wampler have documented the diffusion of participatory budgeting from Porto Alegre to municipalities across diverse political and social contexts. In Brazil, PB has been part of a broader democratization process since the 1980s, which also included the development of health councils, policy conferences, and other participatory forums.

A consistent finding in the literature is that political will and civil society mobilization are essential to the sustainability of PB. Where local governments actively supported the PB and civil society organizations involved in outreach, participation levels tended to be higher and more consistent between administrations. In contrast, a change in political leadership often led to the weakening or discontinuation of PB processes, highlighting the fragility of participatory institutions in certain institutional settings.

Quantitative data underscore both the scale and the variability of PB in Brazil. In approximately 2010, more than 120 Brazilian cities had active PB programs, including major urban centers such as S\~ao Paulo, Belo Horizonte, and Recife. According to the Brennan Center, almost half of the 250 largest Brazilian cities had adopted PB within the first two decades of its existence~\cite{brennancenter1}. Participation was substantial: In Porto Alegre, for example, PB attracted up to 40,000 participants per year in the late 1990s~\cite{archiveepa}.

However, these participants were not always representative of the larger population. Studies noted the overrepresentation of community leaders and the underrepresentation of the most marginalized populations~\cite{archiveepa}. However, even with these limitations, the PB in Brazil succeeded in raising the voice of lower-income residents to a greater extent than traditional political channels~\cite{archiveepa}.

The academic literature on online participation in Brazil is more recent, but is rapidly growing. Brazil has pioneered several digital democracy initiatives. The federal government launched \textit{Participa.br} in the mid-2010s to host national policy consultations, while states such as Rio Grande do Sul implemented platforms like \textit{Gabinete Digital} to crowdsource ideas from the public. Cities such as Belo Horizonte and S\~ao Paulo also launched online PB portals and digital consultation platforms.

These efforts have been accompanied by both enthusiasm and caution. A 2023 report by People Powered on Brazil’s national planning participatory process for the \textit{Pluriannual Plan} (PPA) noted that the government surpassed its goal of engaging 1 million citizens, ultimately involving approximately 1.5 million people through the digital platform~\cite{veciana2023}. However, the success was attributed to a hybrid strategy that combined digital tools with offline outreach and mobilization efforts~\cite{veciana2023}.

Brazilian researchers have also investigated the deliberative quality of online participation. For example, an analysis of the Belo Horizonte digital PB forum by Maia et al. concluded that even when online discussions lacked reciprocal dialogue, they still facilitated the expression of reasoned arguments and the exposure to the viewpoints of others, key ingredients for political learning~\cite{sampaio2011}.

Another important thread in Brazilian scholarship relates to open government and social innovation. The city of S\~ao Paulo’s adoption of the \textit{Participe+} platform was part of its Open Government Partnership (OGP) commitments~\cite{opengov1}. Case studies from the OGP emphasize the role of civil society co-creation and the use of open source software (such as Consul) in building robust participatory systems~\cite{opengov1}. Preliminary indicators suggest that online platforms have substantially increased participation. In S\~ao Paulo, the launch of \textit{Participe+} in 2020 led to a 500\% increase in participation in the city’s ``Citizen Budget'' process, from 2,097 participants in 2019 to 12,354 in 2020~\cite{opengov1}.

This surge highlights how digital tools can reduce barriers and engage new audiences. The COVID-19 pandemic served as a powerful catalyst in this transition, pushing public institutions to develop remote engagement strategies. \textit{Participe+} allowed for large-scale citizen involvement at a time when in-person gatherings were not possible~\cite{opengov1}.

In summary, the Brazilian literature reveals a rich legacy of participatory institutions and a dynamic process of digital adaptation. Although online PB can dramatically expand input, it also presents challenges: How to process large volumes of proposals and ensure equitable representation. As international experiences show, without the response of government actors, even high levels of participation can result in disillusionment or tokenism~\cite{berliner2024}. Brazil’s case illustrates both the potential and the limitations of participatory governance and sets the stage for exploring how emerging technologies like generative AI could strengthen the impact and inclusiveness of participatory budgeting.

\section{Case Study: The Participe+ Platform and Participatory Budgeting in S\~ao Paulo}

\subsection{Background of Participe+ (S\~ao Paulo)}

\textit{Participe+} is the official digital participation platform of the S\~ao Paulo City Hall, launched in July 2020 during the COVID-19 pandemic~\cite{opengov1}. Its creation was accelerated by the need to ensure continuity of public participation during social distancing measures, but was also grounded in years of planning by the Open Government Committee of the city.

The platform was developed through a multi-stakeholder collaboration that involved municipal officials, civil society organizations, academic experts, and the Open Government Partnership (OGP) community~\cite{opengov1}. It is based on \textit{Consul}, an open-source platform originally created by the city of Madrid, and was adapted by S\~ao Paulo to meet local needs~\cite{opengov1}. This adaptation included custom features such as new polling formats and thematic structuring of consultation content.

The adoption of an open-source platform reflects a broader trend of inter-city collaboration and digital democratic innovation. Just as Madrid’s \textit{Decide Madrid} inspired Participe+, the S\~ao Paulo experience may serve as a model for other Brazilian municipalities. From its inception, \textit{Participe+} was envisioned as a comprehensive digital portal, a 'publicly available catalog of online participation tools', capable of facilitating citizen participation in policymaking even during crises.

\textit{Participe+} supports a variety of participation formats, including public consultations (where citizens can comment on draft policies or plans), participatory budgeting processes, online polling, and even elections for citizen representatives to councils such as the city’s Multi-Stakeholder Forum. This integrated structure reflects the emergence of centralized civic-tech platforms in large cities with the aim of simplifying digital participation.

\subsection{Participe+ in Participatory Budgeting}

One of the flagship uses of \textit{Participe+} has been its role in S\~ao Paulo’s participatory budgeting (PB), also known as 'Citizen budgeting'. In late 2020, the platform allowed residents across the city's vast urban area to submit spending proposals and vote on priorities entirely online. This led to a reported 500\% increase in participation compared to in-person PB meetings~\cite{opengov1}—from around 2,100 participants in 2019 to more than 12,300 participants in 2020.

This dramatic increase suggests that the online format effectively removed key barriers to participation, such as long travel times and scheduling conflicts, which often deter residents in large cities from attending physical meetings. The digital format likely also attracted a younger and more diverse demographic, as online platforms tend to lower the age and accessibility barriers seen in traditional participatory spaces.

Users of \textit{Participe+} can submit structured PB proposals after logging into Brazil's national government ID system, ensuring one-person-one-account validation. Citizens fill out a form describing a proposed project or spending idea that others can view, comment on, and support. The city manages the PB process through different phases, e.g., idea submission, refinement (including merging similar proposals), and voting. Transparency is enhanced by publishing all proposals and their support metrics, as well as final outcomes and funded projects.

One notable achievement of the platform was its success in engaging previously underrepresented areas. S\~ao Paulo is marked by high geographical and social inequalities; Peripheral neighborhoods often face barriers to participation due to distance from administrative centers. By offering mobile and computer access, \textit{Participe+} lowered these barriers. According to municipal reports, the platform engaged more than 100,000 visitors and registered more than 20,000 participants in its first six months~\cite{opengov1}.

In addition to PB, the platform has hosted significant consultations, including on the city’s Sustainable Development Goals (SDGs) for 2030 and a green space policy plan. These processes attracted thousands of participants, demonstrating the versatility of the platform. Transparency is a key strength: the results of all participatory processes are available in open data formats, allowing civil society and researchers to analyze the results~\cite{opengov1}.

\subsection{Analysis of Participe+ Implementation}

The initial success of \textit{Participe+} can be attributed to several enabling factors:

\begin{enumerate}
  \item \textbf{Timing and urgency:} The pandemic created an immediate need for remote engagement tools, generating both political support and citizen demand.
  \item \textbf{Collaborative development:} Civil society participation and international best practices (including the use of Consul software) contributed to a robust and user-centered design~\cite{opengov1}.
  \item \textbf{Outreach and trust:} As part of the Open Government initiative, the platform benefitted from community mobilization and credibility. Integration with Brazil's secure login system added more trust and ensured the integrity of the vote~\cite{veciana2023}.
  \item \textbf{Usability:} The interface was accessible and intuitive, allowing users to navigate by theme or process and clearly see how to contribute.
\end{enumerate}

Despite its strengths, challenges remain. Internet access and digital literacy are not universal in S\~ao Paulo. The OGP reports note that educational inequalities and lack of access remain obstacles to full inclusion~\cite{opengov1}. To avoid reinforcing inequalities, the city has recognized the need for hybrid models (e.g., access points in community centers) and continued outreach.

Sustained participation is also an issue. Although necessity drove early participation, long-term participation depends on demonstrating impact. Citizens must see that their contributions lead to real outcomes. For this reason, city officials emphasized the importance of expanding the use of the platform and providing regular input opportunities~\cite{opengov1}.

In this context, generative AI tools may offer support. By helping citizens formulate proposals or providing immediate feedback, AI could make participation more accessible, personalized and satisfying, potentially increasing engagement rates and inclusion.

The \textit{Participe+} platform represents a compelling example of how traditional participatory institutions such as PB assemblies can be restructured into digital formats. It demonstrates the transformative potential of civic technology to increase participation, improve transparency, and expand inclusivity. At the same time, the case reveals challenges: digital divides, the risk of declining engagement, and the need for continuous improvement.


\section{Generative AI in Proposal Development for Participatory Budgeting}

One of the most critical stages of participatory budgeting is the proposal development phase, where the initial ideas of the citizens are transformed into concrete, actionable projects. Traditionally, this step is facilitated by in-person support from municipal staff or community experts who help translate general suggestions (e.g., 'improve traffic safety') into detailed project proposals (e.g., 'install three streetlights and a pedestrian crossing at the intersection of X, costing Y.

However, the scalability of human facilitation is limited. In many PB processes, a small number of facilitators must assist dozens or hundreds of participants, leading to uneven proposal quality. Many citizens end up drafting proposals on their own, often with limited writing experience or policy knowledge. This is where generative AI has transformative potential, providing personalized, on-demand assistance throughout the proposal development process.

\subsection{Structured Guidance}

Generative AI can simulate the support of a human facilitator by breaking down the proposal process into guided questions. An AI chatbot, for example, could ask: 'What problem do you want to solve?', 'Who would benefit?', 'What resources are needed?' and 'What is the expected impact?'

The AI can then use these responses to generate a structured draft proposal, ensuring the inclusion of key components: problem statement, solution, beneficiaries, estimated cost, and location. This scaffold approach helps participants organize their thoughts, especially those who may lack confidence or experience in formal writing.

Research has shown that proposal quality in PB is often inconsistent; some submissions are detailed, while others are just one-line ideas. Generative AI can help elevate the baseline by ensuring that all proposals follow a coherent structure and contain essential elements.

\subsection{Enhancing Clarity and Coherence}

Generative AI can also act as a writing assistant, refining user input to enhance clarity. For example, a citizen might write: “The street in front of school is dark and dangerous.” The AI could transform this into: “Install additional lighting on 5th Avenue near Lincoln School to improve safety for students and pedestrians.”

AI can also suggest details: 'Place four new street lamps and paint the sidewalks, similar to recent improvements in District 3'. By improving grammar, specificity, and readability, AI helps make proposals more actionable and easier for other citizens, and city officials, to evaluate.

\subsection{Providing Data and References}

When integrated with open data systems, AI can enrich proposals with contextual information. For example, if the city’s traffic data indicates a history of accidents on a specific road, AI could add: According to city data, three pedestrian accidents occurred on this street in the past year~\cite{brennancenter1}.”

Generative AI can also help estimate costs by referencing previous PB projects. This contributes to more realistic evidence-based proposals and helps bridge the knowledge gap that often exists between technically proficient and less experienced participants~\cite{royalsociety2024}.

\subsection{Encouraging Idea Diversity}

Another application of AI is to prompt a wider range of ideas. Instead of citizens only proposing familiar or obvious projects (for example, 'fix potholes'), AI could suggest alternatives based on topic interests: 'Residents in other districts have proposed community gardens or outdoor fitness equipment; would something like that interest you?'

This brainstorming function helps expose users to new possibilities while still prioritizing their autonomy. AI should act as a creative partner rather than a menu of fixed options, encouraging originality while avoiding homogenization.

\subsection{Refinement and Validation}

After a proposal is drafted, AI can assess its scope and compliance with PB rules. If, for example, a proposal exceeds budget limits or falls outside of municipal jurisdiction, the AI can flag this and suggest revisions.

AI could also detect duplicate proposals and recommend consolidation, for instance, alerting users if five proposals about the same park have been submitted. This improves efficiency and can foster collaboration.

Some advanced applications may include estimating the number of beneficiaries or the likelihood of implementation success based on historical trends. Although such predictions are not precise, they provide useful feedback and stimulate reflection.

\subsection{User Scenario Illustration}

To illustrate the potential, imagine Ana, a resident who wants to propose a computer lab at her local community center:

\begin{quote}
\textbf{Ana:} “Computer lab for kids.”\\
\textbf{AI:} “What problem are you trying to solve?”\\
\textbf{Ana:} “Many students in my neighborhood can’t do homework that needs the internet.”\\
\textbf{AI:} “Where would the lab be and what would it need?”\\
\textbf{Ana:} “In the community center, maybe 10 computers and Wi-Fi.”\\
\textbf{AI:} “Proposal draft: Create a Community Computer Lab at XYZ Center. Description: Equip the XYZ Community Center with a computer lab for students who lack internet access. Includes 10 computers, broadband connection, and staff volunteer training. Estimated cost: \$15,000. Expected beneficiaries: 200+ students and other residents.”\\
\textbf{AI:} “Similar proposal found: ‘Free Wi-Fi at Community Center (by John)' - 70\% overlap. You may want to combine efforts.”
\end{quote}

With AI support, Ana submits a comprehensive proposal, improving the clarity of her contribution and increasing its chances of success. Importantly, the AI enables her idea without taking over; it supports, but Ana remains the author.

\subsection{Democratic and Inclusive Impact}

AI assistance in proposal writing can reduce participation barriers for citizens with limited literacy, digital skills, or confidence. By leveling the playing field, AI can make PB more inclusive and better reflect the diversity of urban populations~\cite{veciana2023}.

For example, AI can help translate spoken or informal language into formal written proposals. It can also provide multilingual support, crucial in cities like S\~ao Paulo with diverse communities.

This aligns with the principles of democratic innovation: expanding access, reducing inequality, and empowering all voices to participate meaningfully.

Generative AI can act as a scalable, intelligent facilitator for participatory budgeting. By offering structured guidance, clarity enhancement, data integration, and inclusive support, it empowers citizens to submit stronger and more impactful proposals.

Beyond improving individual submissions, AI-supported drafting can improve the overall quality of deliberation, helping citizens engage more deeply with each other’s ideas and fostering a culture of thoughtful civic dialogue.

This sets the stage for the next section, where we present a framework for integrating generative AI into the participatory budgeting cycle.

\section{Discussion and Conclusion}

The exploration of generative AI as a catalyst for democratic innovation in participatory budgeting reveals a nuanced but ultimately hopeful picture. Our analysis, centered on the Brazilian context and the case of S\~ao Paulo’s \textit{Participe+} platform, suggests that integrating AI into participatory processes constitutes a meaningful democratic innovation. It builds on the legacy of participatory budgeting, contributes to the reconstruction of participatory practices in the digital age, and may transform how citizens and governments interact in policymaking.

Participatory budgeting itself was a groundbreaking innovation, shifting power by giving citizens direct control over portions of public budgets. The addition of generative AI can be understood as an innovation layered on an existing one: enhancing the process without altering its core democratic function.

If democratic innovation is defined by its ability to increase inclusion, deliberative quality, and citizen influence, then AI-assisted PB qualifies. Enhances inclusiveness by lowering barriers to participation and improves deliberation by helping citizens express structured and informed proposals. Although it does not increase formal power, it strengthens the impact of that power by improving the quality of the input.


The benefits for engagement and democratic legitimacy are potentially substantial. A participatory budgeting process that reaches tens of thousands of citizens, each feeling their input was considered, can significantly rebuild trust between the public and government. Tangible results - public works created from citizen ideas, and inclusive access can increase political efficacy and reduce citizen apathy.

This aligns with the larger democratic aspirations of legacy, reconstruction, and transformation. The legacy of PB is being rebuilt with digital tools and possibly transformed through AI, redefining citizen-government interactions.

However, significant risks remain. Algorithmic bias and access inequality can lead to unintended consequences~\cite{archiveepa, Machado2025}. For example, more wealthy and educated citizens might use AI tools more effectively, dominating proposal spaces, and undermining the equity goals of PB. Therefore, strong equity measures must accompany AI integration: targeted training, support for less digitally literate communities, and monitoring of who is participating and succeeding in the process.

As the \textit{Participe+} experience shows, the digital divide already presents challenges. Adding AI introduces another layer of potential inequality, between those comfortable with AI and those unfamiliar. To address this, hybrid approaches are recommended: Off-line participation options must remain available, and AI should supplement, not replace, human touchpoints.

The responsible deployment of AI in PB requires ethical guidelines and potentially new policy frameworks. Key considerations include algorithmic transparency and accountability mechanisms. For example, publishing aggregate data on AI-generated suggestions and their adoption rates can support public audits and help identify biases.

Legal clarity on data usage is also essential. Municipalities must ensure compliance with data protection laws. In parallel, governments may consider updating civic participation regulations to formally allow AI-assisted input to be considered in decision making, thus legitimizing its role.

Pilot programs will be vital for evidence-based implementation. Small-scale trials in specific districts or thematic areas can yield empirical data on participation rates, user satisfaction, and proposal outcomes. These trials can inform broader rollout strategies.

As AI technologies evolve, continuous updates to the platform and user training will be needed. Cities may benefit from international collaboration, sharing AI tools and practices tailored for PB, similar to how open-source platforms like Consul and Decidim spread globally. This collaborative model could accelerate innovation while avoiding redundant effort.

Beyond PB, the lessons of AI integration could inform other participatory institutions, such as citizen assemblies or collaborative policy design. Generative AI could become a common enabler of richer and more inclusive democratic engagement.

When thoughtfully implemented, generative AI has the potential to enhance participatory budgeting by empowering citizens and streamlined democratic processes. It represents a convergence of technological and democratic innovation.

In an era where democracies seek to reconstruct trust and legitimacy, such local-level innovations may cumulatively foster more participatory and empowered politics. Although technical and ethical challenges lie ahead, the goal, a more inclusive democracy, is well worth the effort.

\newpage

\end{document}